\documentclass[twocolumn,
superscriptaddress,
showpacs,preprintnumbers,
 amsmath,amssymb,
 aps,
prb,
]{revtex4-1}

\usepackage{graphicx}
\usepackage{dcolumn}
\usepackage{bm}

\begin{document}

\preprint{APS}

\title{Structural anomalies and short-range magnetic correlations in the orbitally degenerated system Sr$_2$VO$_4$}

\author{Ichihiro Yamauchi}\email{ichihiro@post.kek.jp}
\affiliation{Muon Science Laboratory and Condensed Matter Research Center, Institute of Materials Structure Science, High Energy Accelerator Research Organization (KEK), Tsukuba, Ibaraki 305-0801, Japan}
\author{Kazuhiro Nawa}
\affiliation{Graduate School of Science, Kyoto University, Kyoto 606-8502, Japan}
\affiliation{Institute for Solid State Physics, University of Tokyo, Kashiwa 277-8581, Japan}
\author{Masatoshi Hiraishi}
\affiliation{Muon Science Laboratory and Condensed Matter Research Center, Institute of Materials Structure Science, High Energy Accelerator Research Organization (KEK), Tsukuba, Ibaraki 305-0801, Japan}
\author{Masanori Miyazaki}
\affiliation{Muon Science Laboratory and Condensed Matter Research Center, Institute of Materials Structure Science, High Energy Accelerator Research Organization (KEK), Tsukuba, Ibaraki 305-0801, Japan}
\author{Akihiro Koda}
\affiliation{Muon Science Laboratory and Condensed Matter Research Center, Institute of Materials Structure Science, High Energy Accelerator Research Organization (KEK), Tsukuba, Ibaraki 305-0801, Japan}
\affiliation{The Graduate University for Advanced Studies (Sokendai), Tsukuba, Ibaraki 305-0801, Japan}
\author{Kenji M. Kojima}
\affiliation{Muon Science Laboratory and Condensed Matter Research Center, Institute of Materials Structure Science, High Energy Accelerator Research Organization (KEK), Tsukuba, Ibaraki 305-0801, Japan}
\affiliation{The Graduate University for Advanced Studies (Sokendai), Tsukuba, Ibaraki 305-0801, Japan}
\author{Ryosuke Kadono}
\affiliation{Muon Science Laboratory and Condensed Matter Research Center, Institute of Materials Structure Science, High Energy Accelerator Research Organization (KEK), Tsukuba, Ibaraki 305-0801, Japan}
\affiliation{The Graduate University for Advanced Studies (Sokendai), Tsukuba, Ibaraki 305-0801, Japan}
\author{Hironori Nakao}
\affiliation{The Graduate University for Advanced Studies (Sokendai), Tsukuba, Ibaraki 305-0801, Japan}
\affiliation{Photon Factory and Condensed Matter Research Center, Institute of Materials Structure Science, High Energy Accelerator Research Organization (KEK), Tsukuba, Ibaraki 305-0801, Japan}
\author{Reiji Kumai}
\affiliation{The Graduate University for Advanced Studies (Sokendai), Tsukuba, Ibaraki 305-0801, Japan}
\affiliation{Photon Factory and Condensed Matter Research Center, Institute of Materials Structure Science, High Energy Accelerator Research Organization (KEK), Tsukuba, Ibaraki 305-0801, Japan}
\author{Youichi Murakami}
\affiliation{The Graduate University for Advanced Studies (Sokendai), Tsukuba, Ibaraki 305-0801, Japan}
\affiliation{Photon Factory and Condensed Matter Research Center, Institute of Materials Structure Science, High Energy Accelerator Research Organization (KEK), Tsukuba, Ibaraki 305-0801, Japan}
\author{Hiroaki Ueda}
\affiliation{Graduate School of Science, Kyoto University, Kyoto 606-8502, Japan}
\author{Kazuyoshi Yoshimura}
\affiliation{Graduate School of Science, Kyoto University, Kyoto 606-8502, Japan}
\author{Masashi Takigawa}
\affiliation{Institute for Solid State Physics, University of Tokyo, Kashiwa 277-8581, Japan}

\date{\today}

\begin{abstract}

We report on the electronic ground state of a layered perovskite vanadium oxide Sr$_2$VO$_4$ studied by the combined use of synchrotron radiation x-ray diffraction (SR-XRD) and muon spin rotation/relaxation ($\mu$SR) techniques, where $\mu$SR measurements were extended down to 30 mK. We found an intermediate orthorhombic phase between $T_{\rm c2} \sim$~130 K and $T_{\rm c1} \sim$~100 K, whereas a tetragonal phase appears for $T > T_{\rm c2}$ and $T < T_{\rm c1}$. The absence of long-range magnetic order was confirmed by $\mu$SR at the reentrant tetragonal phase below $T_{\rm c1}$, where the relative enhancement in the $c$-axis length versus that of the $a$-axis length was observed. However, no clear indication of the lowering of the tetragonal lattice symmetry with superlattice modulation, which is expected in the orbital order state with superstructure of $d_{yz}$ and $d_{zx}$ orbitals, was observed by SR-XRD below $T_{\rm c1}$. Instead, it was inferred from $\mu$SR that a magnetic state developed below $T_{\rm c0} \sim$~10 K, which was characterized by the highly inhomogeneous and fluctuating local magnetic fields down to 30 mK. We argue that the anomalous magnetic ground state below $T_{\rm c0}$  originates from the coexistence of ferromagnetic and antiferromagnetic correlations.

\end{abstract}

\maketitle


\section{Introduction}
The strong spin-orbit interaction is one of the current topics in condensed matter physics, as it accompanies fascinating phenomena such as the multipole order in $f$-electron systems\cite{Santini2009} and the $J_{\rm eff}=1/2$ Mott transition in 5$d$ electron systems.\cite{Kim2009} In this context, the layered perovskite vanadium oxide Sr$_2$VO$_4$ is attracting renewed interest as a candidate compound for a $d^1$ Mott insulator showing novel spin-orbital magnetism. It has a tetragonal K$_2$NiF$_4$ structure with the space group $I4/mmm$ at room temperature.\cite{Nozaki1991,Zhou2007} The magnetic vanadium ion (V$^{4+}$, 3$d^1$) is octahedrally coordinated with six ligand oxygen ions, where the octahedron is elongated along the $c$ axis. The anisotropy of the octahedron leads to a tetragonal crystal field that partially lifts the three-fold degeneracy of the $t_{2g}$ orbitals by splitting them into $d_{xy}$ and $d_{yz}/d_{zx}$ orbitals with $d_{xy}$ having a higher energy. Thus, the single $d$ electron would occupy the unquenched $d_{yz}/d_{zx}$ orbitals in case of a negligible spin-orbit interaction.\cite{Zhou2007}

However, it has been pointed out that the spin-orbit coupling may be large enough to split the $d_{yz}/d_{zx}$ state into two Kramers doublets.\cite{Jackeli2009} The Kramers doublets are mapped to isospin up and down states, which consist of the electron spin and orbital angular momentum. It was also suggested that a magnetic octupole order (an antiferroic order of the isospins) would be the ground state of Sr$_2$VO$_4$. Furthermore, an alternating scenario of a spin-orbital order with almost muted magnetic moments has been proposed for the ground state of this compound.\cite{Eremin2011}

Despite the efforts of several groups to synthesize high-quality powdered samples in search for signatures of novel spin-orbital magnetism, the varying electronic properties reported in earlier studies seem to call for precautions in assessing the results in relation to the sample quality. The magnetic susceptibility ($\chi$) and heat capacity measurements provided a common ground that the compound successively undergoes three phase transitions at $T_{\rm c0}\sim$~10 K, $T_{\rm c1}\sim$~100 K, and $T_{\rm c2} \sim$~130 K.\cite{Zhou2007,Viennois2010,Teyssier2011} A report on an in-house x-ray diffraction (XRD) experiment showed the splittings of several reflection peaks between $T_{\rm c1}$ and $T_{\rm c2}$.\cite{Zhou2007} It was also reported that a rapid enhancement in the $c/a$ ratio (where $c$ and $a$ are the lengths of the unit cell along the respective crystalline axes) occurred below $T_{\rm c1}$. \cite{Zhou2007,Katsufuji2014}

Regarding the magnetic properties, a sudden decrease in $\chi$ was observed at $T_{\rm c1}$, although the magnitude of the reduction varied from sample to sample.\cite{Zhou2007,Viennois2010,Nawa2012,Sugiyama2014,Sugiyama2014muSR} Below $T_{\rm c0}$, a clear hysteresis was observed in magnetization measurements, where the magnetism was confirmed to be a bulk property.\cite{Viennois2010,Sugiyama2014,Sugiyama2014muSR} The magnetic hysteresis implies the appearance of weak ferromagnetism below $T_{\rm c0}$. Although no experimental evidence for the magnetic octupole order has been reported so far, recent inelastic neutron scattering experiments suggest that the spin-orbit coupling is strong enough to induce excitation between the crystal field levels.\cite{Zhou2010} Further experimental investigation is required to clarify the microscopic details of magnetism in Sr$_2$VO$_4$.

A previous study on muon spin rotation/relaxation ($\mu$SR) reported a spontaneous internal field at the muon site with a broad distribution and residual magnetic fluctuation in the temperature ($T$) range of 1.8 K $\leq T \leq T_{\rm c0}$.\cite{Sugiyama2014,Sugiyama2014muSR} These results suggest that the spin-orbital ordering state, which was proposed by first-principles calculation studies by Imai \textit{et al.}\cite{Imada2005,Imada2006}, appeared below $T_{\rm c0}$. On the basis of the theoretical studies, we can expect a severe competition between many types of spin-orbital ordering patterns due to the coexistence of ferromagnetic and antiferromagnetic correlations and the associated effect of frustration. In such an exotic ground state, it is expected that the persistent spin dynamics and short-range correlations are found far below the magnetic ordering temperature. Then, we extended the $\mu$SR experiments down to 30 mK.

In this paper, we report on the electronic properties of Sr$_2$VO$_4$ inferred from the combined use of synchrotron radiation x-ray diffraction (SR-XRD) and $\mu$SR measurements of powdered samples of the same batch. From the SR-XRD results, we found an intermediate orthorhombic phase in the $T$ range $T_{\rm c1} \leq T \leq T_{\rm c2}$, whereas a tetragonal phase appears for $T > T_{\rm c2}$ and $T_{\rm c1}$. It is demonstrated by $\mu$SR that the anomalies at $T_{\rm c1}$ and $T_{\rm c2}$ do not accompany long-range magnetic order. Moreover, highly inhomogeneous magnetism is observed to develop below $T_{\rm c0}$. A possible model of local field distribution is introduced to discuss the observed $\mu$SR lineshape and magnetism behind it over the relevant $T$ region.


\section{Experiment}

We prepared two batches of polycrystalline samples (labeled sample-A and B with a net weight of $\sim$300 mg and $\sim$30 mg, respectively) by a two step solid-state reaction.\cite{Nawa2012} First, the precursor Sr$_{10}$(VO$_4$)$_6$(OH)$_2$ was prepared by sintering a stoichiometric mixture of Sr(NO$_3$)$_2$ and V$_2$O$_5$ at 550-800 $^{\circ}$C for 84 h in air. Then, Sr$_{10}$(VO$_4$)$_6$(OH)$_2$ and SrO were mixed at a molar ratio of 1:2 and sintered at 950 $^{\circ}$C for 162 h in a H$_2$ gas flow. $\mu$SR measurements were conducted for sample-A, whereas $\chi$ and SR-XRD measurements were performed for both samples. $\chi$ was measured over a $T$ range between 2 K and 300 K in the presence of a magnetic field of 0.2 T with a superconducting quantum interference device (SQUID).

An SR-XRD experiment was carried out using the imaging-plate diffractometer installed on beamline 8A at the Photon Factory at KEK (KEK-PF). The incident beam was monochromatized by a Si(111) double crystal. X-ray diffraction patterns were measured in the $T$ range of 47-305 K by using a He blower with the X-ray wavelength of 0.077534(3) nm. The lattice constants were obtained by the Le Bail whole-profile fitting to the SR-XRD patterns.\cite{LeBail} The Le Bail whole-profile refinements were performed using the RIETAN-FP software package.\cite{RIETAN}

A conventional $\mu$SR experiment was conducted above 4.2 K using the D1 instrument at J-PARC MUSE and down to 30 mK using the DR spectrometer furnished with a dilution refrigerator on the M15 beamline of TRIUMF, Canada. The muon spin depolarization function $G(t)$ was monitored by measuring the decay-positron asymmetry
\begin{equation}
	A(t) = A_0 G(t) + A_{\rm BG},
\label{eq:A}
\end{equation}
where $A_0$ is the initial asymmetry, and $A_{\rm BG}$ is the background contribution mainly ascribed to the sample holder made of silver (in which the muon depolarization is negligible). The $\mu$SR time spectra $A(t)$ was then analyzed by curve fitting, where we employed the recursive function $G(t)$ derived from the appropriate models for the local field distribution (see the next section for more detail). We obtained $A_{\rm BG}=0.0221(6)$ for J-PARC MUSE D1 and 0.083(3) for TRIUMF M15 DR from the calibration measurements.

\section{Results and discussion}

\subsection{Magnetic susceptibility}

\begin{figure}[h]
\includegraphics[width=7cm,clip]{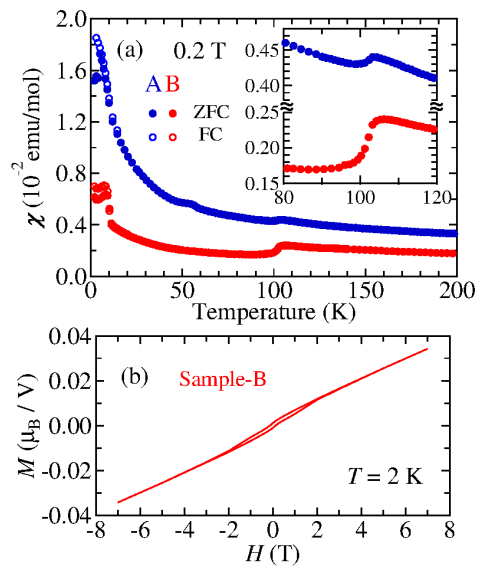}
\caption{\label{fig:1} (Color online) (a) Temperature dependence of the magnetic susceptibility for polycrystalline samples of Sr$_2$VO$_4$, sample-A (blue) and sample-B (red). The filled and empty circles represent data for the zero-field-cooling (ZFC) and field-cooling (FC) processes, respectively. The inset shows a magnified view near $\sim$100 K. (b) A magnetization curve at 2 K for sample-B.}
\end{figure}

Figure \ref{fig:1}(a) shows the $T$ dependence of $\chi$ for two polycrystalline samples used in the present experiment. We observed a sudden decrease in $\chi$ at $T_{\rm c1}$ in both results, despite their quantitative differences. We also observed a clear difference in the $T$ dependence of magnetization between the zero-field-cooling (ZFC) and field-cooling (FC) processes below $T_{\rm c0}$. We can see a clear magnetic hysteresis in the magnetization curve at 2 K in Fig. \ref{fig:1}(b), suggesting weak ferromagnetism. The saturated ferromagnetic moment is estimated to be $\sim3\times10^{-3}\mu_{\rm B}$/V. The data for sample-A exhibit a small cusp at $\sim$55 K in addition to the anomalies at $T_{\rm c1}$ and $T_{\rm c0}$. Although we could not identify the origin of the cusp at present, we note that the peak temperature is slightly higher than the liquid-solid phase transition temperature of oxygen, which is usually observed at $\sim$50 K.

The inset in Fig. \ref{fig:1}(a) shows a magnified view near 100 K, and it clearly shows that sample-B exhibits a larger decrease in $\chi$ at $T_{\rm c1}$ than sample-A. It was reported that the anomalies at $T_{\rm c1}$ and $T_{\rm c0}$ were gradually suppressed as the oxygen deficiency $\delta$ increased for the formula unit (Sr$_2$VO$_{4-\delta}$).\cite{Katsufuji2014} This suggests that sample-A is slightly oxygen off-stoichiometric, although both batches of samples were prepared by the same synthetic method. The oxygen off-stoichiometry may be introduced by the preparation with complex precursor materials and a hydrogen treatment. These facts indicate that we need greater control of the synthesis conditions for more precise control of the oxygen content.




\subsection{Synchrotron X-ray diffraction}

\begin{figure}
\includegraphics[width=7cm,clip]{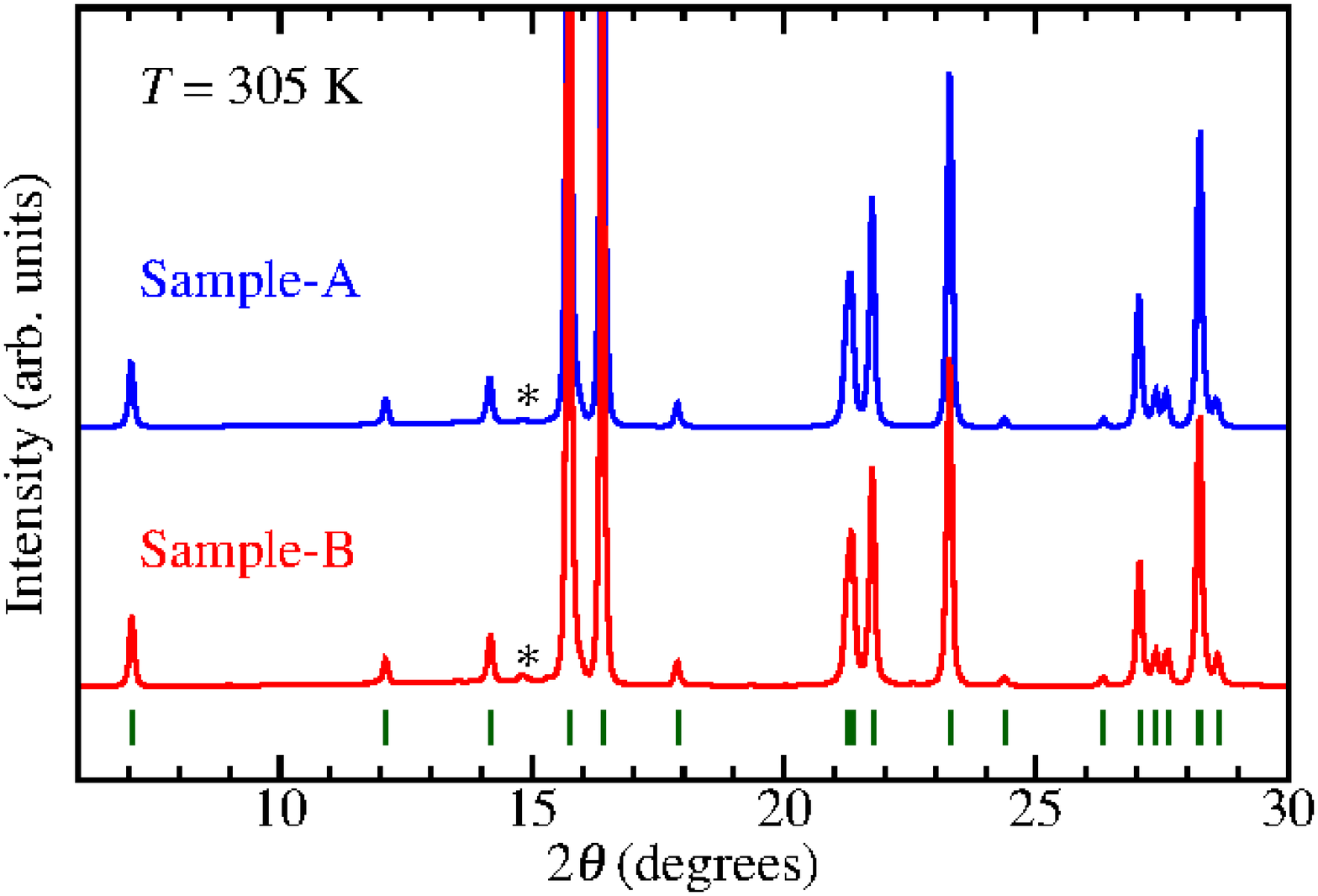}
\caption{\label{fig:2} (Color online) The SR-XRD patterns at 305 K for sample-A (blue) and sample-B (red). The green tick marks indicate the peak positions of the main phase. The asterisks indicate the peaks from the secondary phase. }
\end{figure}

\begin{figure}
\includegraphics[width=7.5cm,clip]{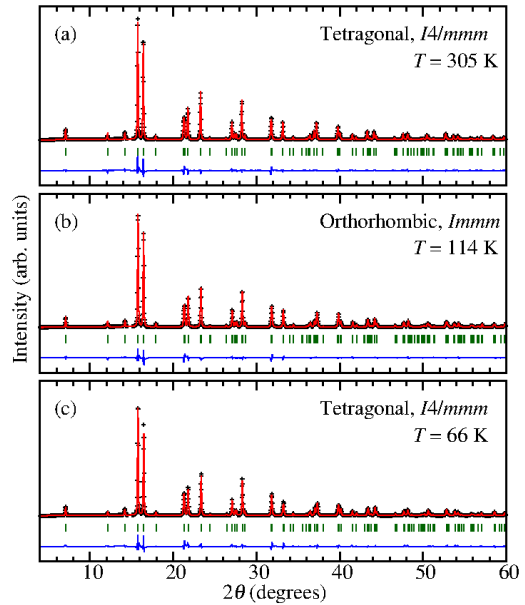}
\caption{\label{fig:3} (Color online) Typical examples of synchrotron x-ray diffraction data measured at (a) 305 K, (b) 114 K, and (c) 66 K. The red lines are the fitted results of the Le Bail whole-profile refinement. The green tick marks represent the positions of the calculated reflections. The blue lines denote the difference between the observed and calculated profiles.}
\end{figure}

\begin{figure}
\includegraphics[width=7cm,clip]{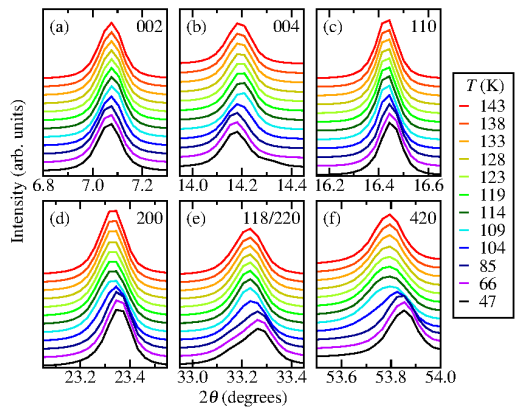}
\caption{\label{fig:4} (Color online) Temperature variation of the selected peaks in the synchrotron radiation x-ray diffraction patterns.}
\end{figure}

The structural properties of the present samples were investigated using the SR-XRD technique. Figure \ref{fig:2} shows the SR-XRD patterns at 305 K for sample-A and B. It has been confirmed that the samples were mostly single-phase Sr$_2$VO$_4$ with small amount of a secondary phase at 2$\theta \sim 15^{\circ}$. The secondary phase was identified as Sr$_{10}$(VO$_4$)$_6$(OH)$_2$ used as the precursor during synthesis. Hereafter, we only show the SR-XRD results for sample-B because there are very few differences between the two samples.

Figure \ref{fig:3} shows the diffraction patterns measured at (a) 305 K, (b) 114 K, and (c) 66 K with the results of the Le Bail whole-profile fitting method. The SR-XRD pattern at 305 K was indexed by a tetragonal structure with the $I4/mmm$ space group. The lattice parameters obtained were $a=0.38402(4)$ nm and $c=1.25581(8)$ nm for sample-A and $a=0.38393(6)$ nm and $c=1.25482(8)$ nm for sample-B at 305 K.

\begin{figure}
\includegraphics[width=7cm,clip]{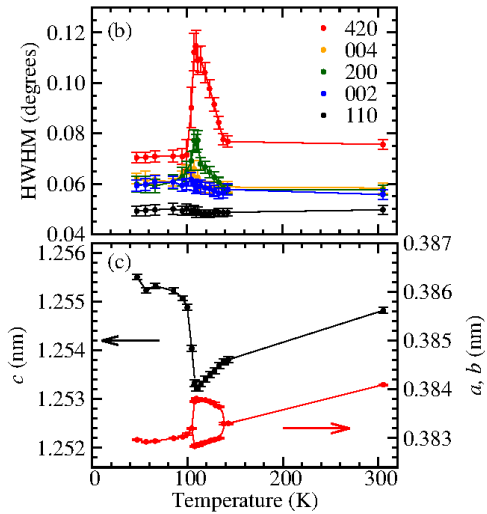}
\caption{\label{fig:5} (Color online) (a) Temperature dependence of the half width at half maximum (HWHM) for selected peaks. (b) Temperature dependence of lattice constants. Solid lines are guide for the eyes.}
\end{figure}

It has been confirmed by the SR-XRD experiment of the present sample that a structural transition occurs at $T_{\rm c2}$ and $T_{\rm c1}$, as shown in previous studies.\cite{Zhou2007,Teyssier2011,Katsufuji2014} Figure \ref{fig:5}(a) shows the $T$ variation of the half width at half maximum (HWHM) for the selected diffraction peaks. Although we could not find any additional peaks in the SR-XRD patterns, the HWHM of the 200 and 420 peaks increased with decreasing temperature for $T_{\rm c1} \leq T \leq T_{\rm c2}$. On the other hand, the 110, 004, and 002 peaks did not exhibit line broadening within the experimental accuracy. These results might be attributed to the existence of an orthorhombic phase for $T_{\rm c1} \leq T \leq T_{\rm c2}$. The absence of broadening in the 110 peak suggests no diagonal distortion along the $\langle 110 \rangle$ direction. It also excludes the coexistence of two tetragonal phases, which had been suggested in a previous study.\cite{Zhou2007} The enhancement in the HWHM for 420(200) peak is explained by the tiny splitting of the peak into the 420(200) and 240(020) peaks due to the expansion and contraction of the two tetragonal $a$ axes.

The continuous line broadening and a lambda-like anomaly in the $T$ dependence of the specific heat\cite{Teyssier2011} strongly suggest that the phase transition at $T_{\rm c2}$ is second-order. Because no superlattice modulation has been discovered below $T_{\rm c2}$, we can provide the possible orthorhombic space group $Immm$ among the subgroups of $I4/mmm$. We confirmed a clear difference between $I4/mmm$ and $Immm$ in the reliability factor ($R_{\rm wp}$) of the Le Bail analysis for $T_{\rm c1} \leq T \leq T_{\rm c2}$, whereas we did not observe a large difference between them for $T > T_{\rm c2}$ and $T < T_{\rm c1}$. For instance, $R_{\rm wp} = 9.5 \%$ for $I4/mmm$ and $R_{\rm wp} = 8.6 \%$ for $Immm$ at 114 K. The $T$ dependencies of the lattice constants are plotted in Fig. \ref{fig:5}(b). We can see the reduction in the $c$ and $a$ lattice constants above $T_{\rm c2}$, and the $a$-axis length splits into two below $T_{\rm c2}$. However, the origins of the phase transitions at $T_{\rm c2}$ are unknown at present. Single-crystal data are highly desirable for further understanding.

Although line broadening was observed for some reflection peaks for $T_{\rm c1} \leq T \leq T_{\rm c2}$ as shown above, we could not identify clear splitting, broadening, or additional peaks in the relevant SR-XRD patterns below $T_{\rm c1}$. The HWHMs above $T_{\rm c2}$ and below $T_{\rm c1}$ are identical within the experimental uncertainty, as shown in Fig. \ref{fig:5}(a). Furthermore, we obtained similar values of $R_{\rm wp}$ for the Le Bail analysis with the $I4/mmm$ and $Immm$ structures; as shown for $T > T_{\rm c2}$, $R_{\rm wp} = 9.7 \%$ for $I4/mmm$ and $R_{\rm wp} = 9.4 \%$ for $Immm$ at 66 K. These results suggest that a tetragonal $I4/mmm$ phase appears below $T_{\rm c1}$.

Figures \ref{fig:4}(a) and (b) show the 002 and 004 peaks in the XRD patterns observed at various values of $T$. We found that the 002 and 004 peaks shifted to lower angles below $\sim$105 K, whereas the 200, 110, and 420 peaks moved to higher angles, as shown in Figs. \ref{fig:4}(c), (d), and (f). As shown in Fig. \ref{fig:5} (b), these results correspond to a steep enhancement in the $c$-axis length, whereas the $a$-axis length slightly decreases below $T_{\rm c1}$. This explains the asymmetric broadening of the 118 and 220 peaks below $T_{\rm c1}$, as shown in Fig. \ref{fig:4}(e) because these two peaks move in opposite directions owing to the structural change.


Although there are some anomalies in the structural properties, as shown above, we could not find evidence of the lowering of the lattice symmetry and superlattice modulation below $T_{\rm c1}$, which are expected in the orbital order state with the superstructure of the $d_{yz}$ and $d_{zx}$ orbitals. This indicates that the degeneracy of the $d_{\rm yz}/d_{\rm zx}$ orbitals is not lifted below $T_{\rm c1}$. These structural properties below $T_{\rm c1}$ are fully consistent with the previous results reported by Zhou \textit{et al}.\cite{Zhou2007} They suggested that the enhancement in the $c/a$ ratio without orthorhombic distortion might originate from the increased occupancy of the $d_{yz}/d_{zx}$ orbitals in place of the $d_{xy}$ orbital. This seems reasonable because the crystal field splitting between the $d_{xy}$ and $d_{yz}/d_{zx}$ orbitals is quite small ($\sim$0.08 eV in the local density approximation calculation).\cite{Imada2005,Imada2006} Thus, we suggest that the $d_{xy}$ orbital contribution to the electronic state is not negligible in the high-$T$ tetragonal phase above $T_{\rm c2}$ owing to the narrow gap between the $d_{yz}/d_{zx}$ and $d_{xy}$ states. Then, the electronic state changes to the pure $d_{yz}/d_{zx}$ state in the low-$T$ tetragonal phase below $T_{\rm c1}$ owing to an enhancement in the energy gap, which is caused by the increase in the $c/a$ ratio to gain crystal field energy.


\subsection{Muon spin relaxation}

\subsubsection{Anomalous paramagnetism ($T>T_{c0}$)}

In the paramagnetic state of ferromagnetic and antiferromagnetic systems, the muon spin depolarization at zero external field has contributions from the static nuclear dipolar fields and the fluctuating fields from electron spins. The nuclear spin contribution is well reproduced by the Gaussian Kubo-Toyabe (GKT) function derived from the Gaussian distribution of the local field ($B_{\rm loc}$), which is a good approximation for a dense ensemble of randomly orientated nuclear magnetic moments.\cite{Hayano_KT} The GKT function is written as
\begin{equation}
	G_{\rm GKT}(t) = \frac{1}{3} + \frac{2}{3} \left[1-(\Delta t)^2\right] \exp\left[-\frac{(\Delta t)^2}{2}\right],
\label{eq:GKT}
\end{equation}
where $\Delta/\gamma_{\mu} = \langle B_{\rm loc}^2 \rangle ^{1/2}$ corresponds to the root-mean-square for the distribution of the local magnetic field (where $\gamma_{\mu} = 2 \pi \times 135.53$ MHz/T is the muon gyromagnetic ratio). On the other hand, the dynamical electron spin contribution is given as $\exp(-\lambda t)$ with the relaxation rate $\lambda$. Then, the depolarization function $G(t)$ is described by the product of the two contributions:
\begin{equation}
	G(t) = G_{\rm GKT}(t) \exp(-\lambda t).
\label{eq:PM_G}
\end{equation}

\begin{figure}
\includegraphics[width=7cm,clip]{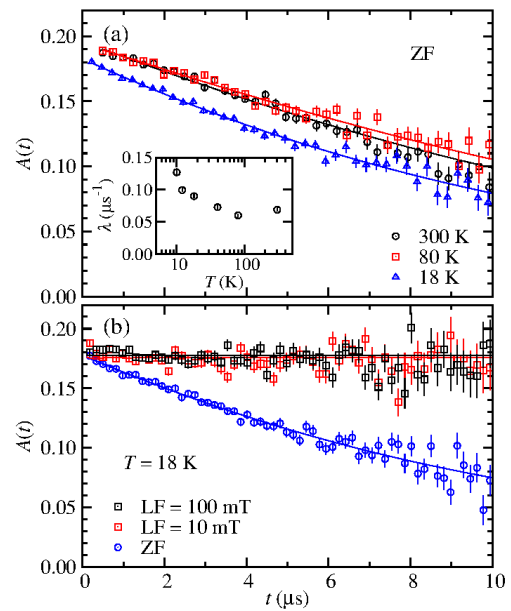}
\caption{\label{fig:6} (Color online) (a) Zero-field (ZF)-$\mu$SR time spectra at 300 K, 80 K, and 18 K. The solid curves are the fitted results of Eq. (\ref{eq:PM_G}). The inset shows the temperature dependence of the relaxation rate. (b) $\mu$SR spectra at 18 K measured with ZF and longitudinal fields (LF~=~10 mT and 100 mT) with the results of the curve fitting with Eq.(\ref{eq:LKT}). }
\end{figure}

Figure \ref{fig:6}(a) shows the zero-field (ZF)-$\mu$SR spectra at 300 K and 80 K (i.e., above and below $T_{\rm c1}$), where no sign of the emergence of the spontaneous internal magnetic field, such as oscillation and/or fast damping, is observed below $T_{\rm c1}$. This immediately leads us to the conclusion that no long-range magnetic order is present below $T_{c1}$, which is in qualitative agreement with the previous result.\cite{Sugiyama2014} The temperature dependence of $\lambda$ deduced by a curve fitting is plotted in the inset of Fig.~\ref{fig:6}(a).

The spectra in Fig.~\ref{fig:6}(a) including the one at 18 K show that the exponential damping gradually overcomes the GKT depolarization below 40 K. Figure \ref{fig:6}(b) shows $\mu$SR spectra at 18 K measured with ZF and longitudinal fields (LFs) of 10 mT and 100 mT. Generally, the exponential damping due to the electron spin fluctuation does not depend much on the LF because the fluctuation rate of the dynamical field from the ordinary paramagnetic moments is much larger than the Larmor frequency of the muon spin precession. However, the depolarization is quenched by the application of a relatively weak LF, which indicates that $B_{\rm loc}$ is much weaker than the LF and that it is quasistatic within the time window of $\mu$SR ($>10^{-5}$ s). These features allow us to estimate an upper limit for the mean hyperfine field at the muon site, $B_{\rm loc}\simeq\Delta/\gamma_\mu\lesssim$~2 mT, and that of the fluctuation rate $\nu\lesssim$~0.5 MHz. We also performed curve fitting for the LF dependence of the $\mu$SR spectra at 18 K by replacing the exponential term in Eq. (\ref{eq:PM_G}) with the Lorentzian Kubo-Toyabe (LKT) function for a finite LF:
\begin{eqnarray}
	G_{\rm LKT}(t) = && 1-\frac{\Delta}{\omega_0} j_1(\omega_0 t) e^{-\Delta t} \nonumber \\
	 &&  - \bigl( \frac{\Delta}{\omega_0 t} \bigr)^2 [j_0(\omega_0 t) e^{-\Delta t}-1] \nonumber \\
         &&  - \bigl[1 + \bigl(\frac{\Delta}{\omega_0}\bigr)^2 \bigr] \Delta \int^t_0 j_0(\omega_0 t^{\prime}) e^{-\Delta t^{\prime}} d t^{\prime},
\label{eq:LKT}
\end{eqnarray}
where $\omega_0=\gamma_\mu B_0$ is the Larmor frequency of the muon spin precession in the presence of an external field $B_0$, and $j_0$ and $j_1$ denote spherical Bessel functions.\cite{Uemura_SG} We note that the results of the curve fitting indicate that the entire volume of the sample exhibits a Lorentzian-like field distribution. Thus, we are led to the conclusion that the dilute magnetic moments of $d$ electrons having an unusually quasistatic characteristic make a significant contribution to the muon depolarization for $T_{c0} \leq T \leq T_{c1}$, although their origin is not clear at this stage.

\subsubsection{Inhomogeneous magnetism ($T<T_{c0}$)}

\begin{figure}
\includegraphics[width=7cm,clip]{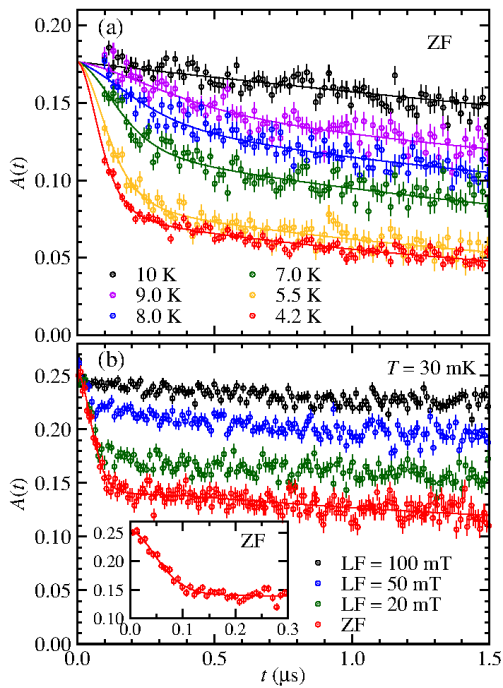}
\caption{\label{fig:7} (Color online) (a) ZF-$\mu$SR time spectra in the temperature range between 4.2 K and 10 K. (b) $\mu$SR time spectra measured at 30 mK in the presence of various longitudinal fields (LFs) up to 100 mT. The inset shows the earlier part of the spectrum at ZF. The solid lines are the fitted results by using Eq. (\ref{eq:G_MO}).}.
\end{figure}

As is observed for the ZF-$\mu$SR spectra in Fig.~\ref{fig:7}(a), a rapid depolarization develops with decreasing temperature below $T_{\rm c0}$. We show the ZF- and LF-$\mu$SR spectra at 30 mK measured in the presence of various LFs up to 100 mT in Fig.~\ref{fig:7}(b). The ZF spectra clearly demonstrate the absence of conventional long-range magnetic order at the base temperature. The initial fast depolarization exhibits a gradual suppression with increasing LF and is nearly quenched at LF~=~100 mT. Surprisingly, we also observed a finite slope due to the residual magnetic fluctuation in these spectra at a large time, which is almost independent of the LF, even at 30 mK. These features can be observed from the fast damped 2/3 amplitude followed by a slowly relaxing 1/3 component originating from the magnetic ordered state with slow persistent spin dynamics.

In a previous $\mu$SR study, Sugiyama $\textit{et al.}$ analyzed their ZF-$\mu$SR spectra by using a combination of an exponentially relaxing cosine oscillation for the damped 2/3 component and an exponential function for the 1/3 tail signal.\cite{Sugiyama2014} They showed that the ratio of the relaxation rate of the 1/3 tail component $\lambda_{\rm tail}$ to the precession frequency $\omega_{\rm AF}$ is close to unity, $\lambda_{\rm tail}/\omega_{\rm AF} \sim 1$, where $\lambda_{\rm tail}/\omega_{\rm AF}$ corresponds to the field distribution normalized to the mean value of $B_{\rm loc}$. However, their fitting function is based on an unique precession signal at a spontaneous local field for the conventional magnetic ordered state, $\lambda_{\rm tail}/\omega_{\rm AF} \ll 1 $ . The observed broad $B_{\rm loc}$ distribution and residual magnetic fluctuation strongly suggest that a highly inhomogeneous magnetic state appears in the ground state of Sr$_2$VO$_4$.

The inset of Fig.~\ref{fig:7}(b) shows the earlier part of the ZF-$\mu$SR spectrum at 30 mK. As mentioned above, the GKT and LKT functions are widely attributed to the $\mu$SR spectra for random magnetism in dense and diluted spin systems. However, the ZF spectrum at 30 mK lacks a decrease that usually precedes the asymptotic tail for the GKT function, although it exhibits a Gaussian-like initial depolarization. We attributed this to a broad distribution of $B_{\rm loc}$, which is not characterized by a single root-mean-square width, and adopted the Gaussian-broadened Gaussian (GbG) function as a recursive function for the curve fitting.\cite{Noakes1997,Noakes1999,Yaouanc2013}

The GbG function is described as one of the generalized forms of the GKT function, where $\Delta$ in Eq.~(\ref{eq:GKT}) has a distribution with the mean value $\Delta_0$ and root-mean-square width $W$. $G_{\rm GbG}(t)$ for ZF is written as
\begin{eqnarray}
	G_{\rm GbG}(t) =&& \frac{1}{3} + \frac{2}{3} \Bigl( \frac{1}{1 + R^2 \Delta_0^2 t^2} \Bigr)^{3/2}   \Bigl( 1-\frac{\Delta_0^2 t^2}{1 + R^2 \Delta_0^2 t^2} \Bigr) \nonumber \\
	&&\times \exp \Bigl( - \frac{\Delta_0^2 t^2}{2(1+R^2 \Delta_0^2 t^2)} \Bigr) ,
\label{eq:GbG}
\end{eqnarray}
where $R$ is the ratio of the distribution width, and $R = W/\Delta_0$. For $R = 0$, the GbG function is identical to the GKT function with a deep minimum after the initial depolarization. The depth of the minimum decreases with increasing $R$ and becomes almost independent of $R$ for $R > 1$. Thus, we imposed the condition $0 \leq R \leq 1$ in the following analysis. We note that the GbG function does not approach the LKT function, even for $R \gg 1$.

\begin{figure}
\includegraphics[width=8.5cm,clip]{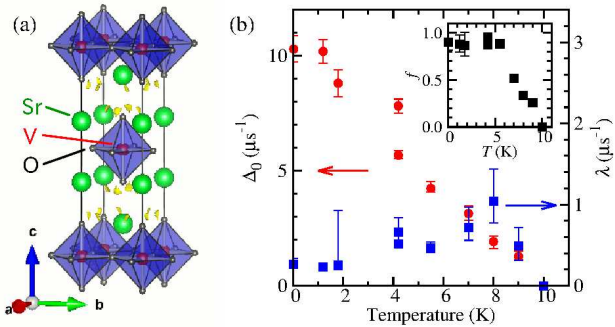}
\caption{\label{fig:8} (Color online) (a) Crystal structure of Sr$_2$VO$_4$ with the minimum regions of the electrostatic potential. The yellow regions define the expected muon site from our Hartree potential calculation by VASP code. \cite{VASP} This figure is drawn by VESTA.\cite{VESTA} (b) Temperature dependencies of $\Delta_0$ (left axis) and the relaxation rate $\lambda$ (right axis). The inset shows the temperature dependence of the volume fraction $f$, showing the inhomogeneous magnetism. }
\end{figure}

For the curve fitting, we presume that $G(t)$ is primarily described by the product of the GbG function and a slow exponential decay due to the persistent spin dynamics in addition to a supplementary term for exponential damping:
\begin{equation}
	G(t) = f G_{{\rm GbG}}(t) e^{-\lambda t} + (1-f) e^{-\lambda_{\rm ext} t},
\label{eq:G_MO}
\end{equation}
where $f$ is the volume fraction exhibiting inhomogeneous magnetism, and $\lambda$ is the relaxation rate reflecting the spin dynamics. The second exponential term with the relaxation rate $\lambda_{\rm ext}$ is for extrinsic contributions originating from the magnetically ordered state region with relatively fast persistent dynamics, which makes the above muon spin depolarization model unreasonable. The $T$ dependence of $f$ is plotted in the inset of Fig.~\ref{fig:8}. The inhomogeneous magnetism develops over the almost entire volume of the sample below 6 K. We obtained $R = 0.7(1)$ below 4.2 K, whereas $R$ is unity at higher temperatures. This discrepancy might be caused by the lack of data for $t \sim$~0.1 $\mu$s that is masked by the limited time resolution of the pulsed muon beam ($\sim$100 ns). We note that the dynamical LKT function and dynamical spin glass models\cite{Uemura_SG} could not reproduce the spectra over the early time range.

The absence of decrease in the $\mu$SR time spectra tells us important aspects of the magnetic ground state. Noakes \textit{et al.} reproduced the shallow decrease in the GbG function by assuming a local field distribution having excess low-field sites.\cite{Noakes1997,Noakes1999}  This implies that the magnetic ordered state is accompanied by a large modulation of the ordered moment magnitude. They also investigated the relationship between $R$ and the correlation length by Monte Carlo simulation. On the basis of their results, $R \sim 0.7$ corresponds to a correlation length of $\sim$4 nm. Imai \textit{et al.} predicted that the spin and orbital correlation functions are reduced as the system size increases.\cite{Imada2005,Imada2006} This result suggests that the correlation length of the spin and orbital degrees of freedom are finitely limited. Our result strongly supports their predictions.

The $T$ dependence of $\Delta_0$ and $\lambda$ are shown in Fig. \ref{fig:8}(b). $\Delta_0$ behaves like an order parameter below $T_{\rm c0}$, approaching 10.3(6) $\mu$s$^{-1}$ at 30 mK. From this value, we can extract a mean field at the muon site, $\Delta_0/\gamma_{\mu} = 12.1(7)$ mT. According to our Hartree potential calculation by using the Vienna ab-initio simulation package (VASP),\cite{VASP} the potential minima for a muon are located around the (0.50, 0.24, 0.69) position, as shown in Fig. \ref{fig:8}(a). Assuming randomly oriented 1$\mu_{\rm B}$ V $d$-electron moments, we obtained $\langle B_{\rm loc}^2\rangle^{1/2}\sim$52 mT/$\mu_{\rm B}$ at the relevant muon site. Thus, we roughly estimate the ordered moment at the V site to be $\sim 0.2 \mu_{\rm B}$. This value is much smaller than the dipole moment for the V$^{4+}$(3$d^1$) ion with quenching of the orbital moment, $\mu_{\rm eff} \sim 1.7\mu_{\rm B}$ for $S=1/2$. On the other hand, $\lambda$ slightly decreases with decreasing $T$ indicating the gradual suppression of the persistent dynamics. However, we observed finite residual magnetic fluctuation even at 30 mK. Persistent spin dynamics have been observed for the magnetic ordered state of several geometrically frustrated magnetic systems. \cite{Dunsiger1996,Hodges2002,Dalmas2006,Dalmas2006PRL,Dally2014}

Finally, we briefly discuss the origin of the inhomogeneous magnetism. As shown in Fig. \ref{fig:1}(a), the magnetization data suggest the existence of the oxygen off-stoichiometry in sample-A, which is used in the present $\mu$SR experiment. However, we could not find any line broadenings and additional peaks in the SR-XRD pattern for sample-A, as shown in Fig. \ref{fig:2}. This indicates that there are no severe deformation and crystallographic randomness due to the oxygen off-stoichiometry in sample-A. From these results, we can consider that the inhomogeneous magnetism might have an intrinsic origin, although we would observe a qualitative difference in the field distribution, correlation length, and relaxation rate depending on the magnitude of the oxygen off-stoichiometry. Imai \textit{et al.} also revealed that there is severe competition between many spin-orbital order patterns due to the coexistence of ferromagnetic and antiferromagnetic interactions and the associated effect of frustration in Sr$_2$VO$_4$.\cite{Imada2005,Imada2006} In their proposed spin-orbital ordering patterns, orthogonal and not strictly orthogonal nearest-neighboring orbital configurations appeared. Such orbital configurations induce the coexistence of ferromagnetic and antiferromagnetic signs of the exchange interactions because of the Goodenough-Kanamori rule. We suggest that the observed broad $B_{\rm loc}$ distribution, short-range correlations, and persistent spin dynamics are signatures of the severe competition between the possible spin-orbital ordered patterns in Sr$_2$VO$_4$.



\section{Summary}

We have conducted SR-XRD and $\mu$SR measurements of polycrystalline samples of Sr$_2$VO$_4$. The SR-XRD results suggest the existence of an intermediate orthorhombic phase between $T_{\rm c2} \sim$~130 K and $T_{\rm c1} \sim$~100 K, whereas a tetragonal phase appears for $T > T_{\rm c2}$ and $T < T_{\rm c1}$. Although a large enhancement in the $c/a$ ratio had been confirmed below $T_{\rm c1}$, we could not find clear evidence for the superlattice modulation and the lowering of the lattice symmetry from tetragonal $I4/mmm$, which is easily expected for the orbital order state with a superlattice structure of $d_{yz}$ and $d_{zx}$ orbitals. These results indicate that the degeneracy of the $d_{yz}/d_{zx}$ orbitals is not lifted, even below $T_{\rm c1}$.

The $\mu$SR data revealed that the structural transition at $T_{\rm c1}$ does not accompany magnetic ordering. On the other hand, we extended the $\mu$SR measurement down to 30 mK and found a highly inhomogeneous magnetic ordered state with slow persistent spin dynamics below $T_{\rm c0} \sim$~10 K. In the magnetic ordered state, the $\mu$SR spectra were well fitted by the GbG function, which is known to describe the muon spin depolarization due to the inhomogeneous magnetism with short-range correlations and a vast moment size distribution. These results suggest that the inhomogeneous magnetic state may be a signature of the severe competition between the many possible spin-orbital ordered states predicted by first-principles calculations \cite{Imada2005,Imada2006}.

\begin{acknowledgments}
The authors thank H. Okabe, S. Takeshita, J. Sugiyama, Y. Kato, and M. Itoh for fruitful discussions. We also thank A. Sagayama, H. Lee, and the staff of TRIUMF and J-PARC MUSE for their technical support.
\end{acknowledgments}


\end{document}